\documentclass[conference]{IEEEtran}
\usepackage{amsmath}
\usepackage[pdftex]{graphicx}
\usepackage{graphicx}
\usepackage{pstricks}
\usepackage{amsfonts}
\usepackage{tabularx}
\newtheorem{Lemma}{Lemma}
\newtheorem{Remark}{Remark}

\usepackage{amssymb}
\usepackage{multirow}
\usepackage{epsfig}
\usepackage{epstopdf}

\usepackage[noadjust]{cite} 
\usepackage{algorithm,algpseudocode,graphicx}

\begin{document}
\title{Decoding Orders and Power Allocation for Untrusted NOMA: A Secrecy Perspective}
\author{\IEEEauthorblockN{Sapna Thapar$^{1}$, Deepak Mishra$^{2}$, and Ravikant Saini$^{1}$}
\IEEEauthorblockA{$^{1}$Department of Electrical Engineering, Indian Institute of Technology Jammu, India\\
$^{2}$School of Electrical Engineering and Telecommunications, University of New South Wales Sydney, Australia \\
Emails: 2018ree0019@iitjammu.ac.in, d.mishra@unsw.edu.au, ravikant.saini@iitjammu.ac.in
}
}
\maketitle

\begin{abstract}
The amalgamation of non-orthogonal multiple access (NOMA) and physical layer security is a significant research interest for providing spectrally-efficient secure fifth-generation networks. Observing the secrecy issue among multiplexed NOMA users, which is stemmed from successive interference cancellation based decoding at receivers, we focus on safeguarding untrusted NOMA. Considering the problem of each user's privacy from each other, the appropriate secure decoding order and power allocation (PA) for users are investigated. Specifically, a decoding order strategy is proposed which is efficient in providing positive secrecy at all NOMA users. An algorithm is also provided through which all the feasible secure decoding orders in accordance with the proposed decoding order strategy can be obtained. Further, in order to maximize the sum secrecy rate of the system, the joint solution of decoding order and PA is obtained numerically. Also, a sub-optimal decoding order solution is proposed. Lastly, numerical results present useful insights on the impact of key system parameters and demonstrate that average secrecy rate performance gain of about $27$ dB is obtained by the jointly optimized solution over different relevant schemes.

\end{abstract}
 

\section{Introduction and Background}
By sharing the same resource block among numerous users \cite{nomasurvey}, non-orthogonal multiple access (NOMA) offers a promising solution to the issue of providing massive connectivity in future wireless networks. However, NOMA is confronted with critical security issues regarding users' privacy, due to the broadcast nature of wireless transmission, and successive interference cancellation (SIC) based decoding at receivers \cite{nomasurvey}-\cite{8114722}. Physical layer security (PLS) is regarded as a favorable technique for providing secure communications over wireless channels \cite{8509094}. Therefore, applying PLS in NOMA has recently emerged as a new research frontier for providing NOMA based spectrally-efficient secure communication system \cite{8114722}. 

\subsection{State-of-the-Art}
The security objectives in NOMA can be divided into two categories: firstly, security of information-carrying signal against external eavesdroppers, and secondly, security of multiplexed data of NOMA users due to breach of trust among users \cite{8509094}. The security concern against external eavesdroppers occurs because the transmitted information is vulnerable to eavesdropping due to the open nature of wireless medium. In this context, PLS of NOMA in large-scale networks has been studied in \cite{liu2017enhancing} where a protected zone around the base station (BS) is adopted to maintain an eavesdropper-free area. In \cite{zhang2016secrecy}, optimal power allocation (PA) is characterized to maximize the secrecy sum rate of multiple users against an eavesdropper in a NOMA system. In \cite{7982788}, the optimal data rates, decoding order, and power assigned to each user is investigated for protecting the data transmission to NOMA users against external eavesdroppers.

Besides, secrecy concern among multiplexed NOMA users is stemmed because of the decoding concept of SIC at receivers. Instead of assuming the users as trusted, secure NOMA transmission among users when the users are considered as untrusted is a practical system design aspect. The untrusted users' scenario is a hostile situation where users have no mutual trust and each user focuses on securing its own data from all others \cite{saini2016ofdma}-\cite{saini2019subcarrier}. In this context, \cite{7833022} assumed near and far users as trusted and untrusted, respectively, in each group of NOMA users and investigated the sum secrecy rate of only near users. In \cite{basepaper} also, the secrecy outage probability of a trusted near user against an untrusted far user is analyzed in a two-user NOMA system. As noted, \cite{7833022}-\cite{basepaper} have assumed only far user as untrusted, however, there exists a serious security risk for the far user also in case the near user is untrusted. Taking this into account, \cite{globecom} has considered a two-user NOMA system assuming both the users as untrusted and proposed a new decoding order that is efficient in providing positive secrecy rate at both the users.

\subsection{Research Gap and Motivation}
On the basis of potential of PLS, \cite{liu2017enhancing}-\cite{7982788} have focused on securing the messages of NOMA users from external eavesdroppers only. In \cite{7833022}-\cite{basepaper}, the secrecy performance of trusted near user is analyzed against untrusted far user in NOMA. Further, \cite{globecom} investigated a new decoding order for untrusted NOMA that is efficient in providing secure data transmission at all users. Note that the work in \cite{globecom} for NOMA security is limited to the study of two users. As inferred, the decoding order selection problem for more users becomes rather complicated because the problem is of combinatorial nature. Therefore, observing the issue of computational complexity, we study the decoding order selection problem for a three-user untrusted NOMA system. However, the proposed investigation can be extended to a general system with more users. Also, keep this in mind that the number of users should not be too large in NOMA due to the practical limitations such as high interference and implementation complexity at receivers with more users \cite{ding2016impact}. Besides, \cite{globecom} assumes perfect SIC that cancels the interference from the decoded users entirely which is a strong assumption due to practical implementation issues \cite{8114722}. Therefore, imperfect SIC model is highly realistic to explore secure NOMA which has been considered in \cite{2019noma} for two-user NOMA case. \textit{To this end, we focus to secure a three-user untrusted NOMA system under practical imperfect SIC based decoding at receivers, which to the best of our knowledge, has not been investigated yet in the literature.} 

\subsection{Key Contributions}
The key contributions of this work are as follows: (1) An appropriate decoding order strategy for a three-user downlink untrusted NOMA system is proposed that safeguards all untrusted users’ confidentiality against each other. (2) Based on the proposed decoding order strategy, an algorithm is provided through which all the feasible secure decoding orders can be obtained. (3) An optimization problem aiming at maximizing the sum secrecy rate of the system is formulated, and the joint solution of decoding order and users' PA is obtained numerically. An efficient sub-optimal decoding order solution is also proposed. (4) Numerical results present useful insights on the optimized solution and analyze secrecy rate performance achieved by the sub-optimal decoding order. 
\section{Secure NOMA among Untrusted Users}\label{section2}
Here we present the system model, proposed decoding order representation, and achievable secrecy rates at NOMA users. 
\subsection{System Model and Downlink NOMA Transmission}
\begin{figure}[!t]
\centering
\includegraphics[scale=.33]{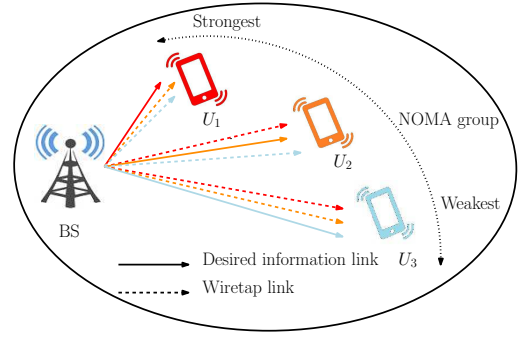}
\caption{Illustration of downlink NOMA system with three untrusted users.}
\label{systemmodel}
\end{figure}

As shown in Fig. \ref{systemmodel}, we consider the downlink NOMA system with a BS and $3$ untrusted users, $U_{n}$, $n\in \mathcal{N}=\{1, 2, 3\}$. All the nodes are equipped with a single antenna. The Rayleigh fading channel gain coefficient between BS and $U_{n}$ is denoted by $h_{n}$. The channel power gains $|h_{n}|^{2}$ follows exponential distribution with mean $\lambda_{n}=L_{p}d_{n}^{-e}$, where $L_{p}$, $e$ and $d_{n}$ denote path loss constant, path loss exponent, and distance from BS to $U_{n}$, respectively. The channel power gains are assumed to be sorted as $|h_{1}|^{2}>|h_{2}|^{2}>|h_{3}|^{2}$. $P_{t}$ denotes the total power transmitted by BS, and $\alpha_{n}$ is PA factor denoting the fraction of $P_{t}$ allocated to $U_{n}$ with $\alpha_{1} + \alpha_{2} + \alpha_{3} = 1$.

In NOMA, the message signals dedicated to the users are superimposed at the BS and then transmitted to the users. The signal transmitted by the BS can be written as
\begin{equation}
x=\sum_{n\in \mathcal{N}}\sqrt{\alpha_{n}P_{t}}x_{n},
\end{equation}
where $x_{n}$ is the unit power signal which contains the message required by $U_{n}$. The signal received by $U_{n}$ can be written as
\begin{equation}
y_{n} = h_{n}x + w_{n},
\end{equation}
where $w_{n}$ is zero-mean additive white Gaussian noise with variance $\sigma^{2}$ at $U_{n}$. After obtaining the received signal, SIC is carried out at receivers to extract the desired message from the multiplexed signal. At each step of SIC, the previously decoded user signals are canceled out from the received signal. Since we consider imperfect SIC model, SIC is not performed ideally, and thus, the residual interference from imperfectly decoded users exists while decoding later users \cite{7881111}.  

\subsection{Decoding Order Representation for Untrusted NOMA}
In NOMA, the order in which the SIC process is completed at users, is represented by a decoding order. In the case of conventional decoding order \cite{nomasurvey}, SIC is applied at trusted nearer users, and the respectively farther users can never decode those nearer users' data. In contrast, untrusted NOMA case considers the practical situation in which all the users are considered as untrusted and they may decode others' information through SIC \cite{basepaper}. Thus, each user follows an independent decoding order, and hence, an overall system decoding order for a $3$-user NOMA system can be represented by a $3\times3$ matrix as shown in Fig. \ref{decoding_order_strategy}. Let us define $o$-th decoding order as $\mathcal{D}_{o}=[{D}_{1}, D_{2}, D_{3}]$, where each column $D_{m}=[D_{1m}, D_{2m}, D_{3m}]^T$, represents the SIC order followed by $U_{m}$. Specifically, $D_{km}=n$ signifies that $U_{m}$ decodes data of $U_{n}$ at $k$-th stage ($k$-th row of matrix), where $k, m, n \in \mathcal{N}$ and $D_{1m} \neq D_{2m} \neq D_{3m}$. Using the concept of permutation, the total possible decoding orders for each user are $3!=6$, such as, $\mathbb{P}=\{P_{z}\}=\{[1,2,3]^T, [1,3,2]^T,[2,1,3]^T, [2,3,1]^T, [3,1,2]^T, [3,2,1]^T\}$, where $z \in \{1, 2...6\}$. Since each user is independent, the count of system decoding orders is $3!\times3!\times3!=216$. Algorithm \ref{Algo:ALa}, given at the top of next page, provides $\mathbb{D} = \{\mathcal{D}_{o}\}$, a set of total possible system decoding orders, where $o \in \{1, 2...216\}$.

\subsection{Achievable Secrecy Rates at Users}
For a decoding order $\mathcal{D}_{o}$, let us denote $\mathbb{S}_{om}^{bn}$ as the set of those users' indices, $i$, which are decoded \textit{before} decoding of $U_{n}$ by $U_{m}$, and $\mathbb{S}_{om}^{an}$ as the set of those users's indices, $j$, which will be decoded \textit{after} decoding of $U_{n}$ by $U_{m}$, where $i, j \in \mathcal{N}\setminus\{n\}$ and $\mathbb{S}_{om}^{bn} \neq \mathbb{S}_{om}^{an}$. Thus, the signal-to-interference-plus-noise-ratio, $\gamma^{o}_{nm}$, when $U_{n}$ is decoded by $U_{m}$, can be defined as \cite{zhang2016secrecy}, \cite{7881111}
\begin{equation}\label{SINR}
\gamma^{o}_{nm}= \frac{\alpha_{n} |h_{m}|^2}{(\zeta\sum\limits_{ i\in \mathbb{S}_{om}^{bn} }\alpha_{i}+\sum\limits_{ j\in \mathbb{S}_{om}^{an}}\alpha_{j})|h_{m}|^2+\frac{1}{\rho_{t}}}, 
\end{equation}
where $\rho_{t}\stackrel{\Delta}{=}P_{t}$/$\sigma^{2}$ is BS transmit signal-to-noise ratio and $\zeta$ is residual interference factor, $(0\leq\zeta\leq1)$, where $\zeta=0$ and $\zeta=1$, respectively, represent perfect SIC and fully imperfect SIC \cite{7881111}. The corresponding data rate can be given as \cite{tse2005fundamentals}
\begin{equation} \label{info_rate}
R^{o}_{nm} = \log_{2}(1+\gamma^{o}_{nm}).
\end{equation}

Let $R^{o}_{sn}$ represents the achievable secrecy rate of $U_{n}$ considering all other users as eavesdroppers. $R^{o}_{sn}$ can be defined as the difference of the rates when $U_{n}$ decodes itself, and the maximum of the rate that other users achieve while decoding data of $U_{n}$ \cite{8509094}. Mathematically, it can be expressed as
\begin{equation}\label{secrecy_rate}
R^{o}_{sn} = \{R^{o}_{nn} - \underset{m\in\mathcal{N}\backslash n}{\max} R^{o}_{nm}\}^+.
\end{equation}

\begin{algorithm}[!t]
{\small
\caption{\small Finding the total possible system decoding orders for a three-user untrusted NOMA system.}\label{Algo:ALa}
\begin{algorithmic}[1]
\Require Set $\mathbb{P}$ \hspace{1.1cm} $\triangleright$ set of possible decoding orders at each user
\Ensure Set $\mathbb{D}$ \hspace{0.85cm} $\triangleright$ set of total possible system decoding orders
\State Set index $o=0$
\For{$u=1$ to $6$}   \hspace{1cm}  
\State $D_{3}= P_{u}$
\For{$v=1$ to $6$} \hspace{1cm} 
\State $D_{2}=P_{v}$
\For{$w=1$ to $6$}   \hspace{1cm} 
\State $o=o+1$
\State $D_{1}=P_{w}$
\State $\mathcal{D}_{o}=[D_{1}, D_{2}, D_{3}]$
\State return $\mathbb{D}=\{\mathcal{D}_{o}\}$
\EndFor
\EndFor
\EndFor
\end{algorithmic}
}
\end{algorithm}

Here the key idea of achieving positive secrecy is to ensure that the rate of the desired channel is higher than that of the eavesdroppers' channel, i.e., the condition $R^{o}_{nn}> R^{o}_{nm}$, simplified as $\gamma^{o}_{nn} > \gamma^{o}_{nm}$ must be satisfied.

\section{Proposed Decoding Order Strategy}
To secure an untrusted NOMA system, we focus on all users' data confidentiality against each other. In this context, we first highlight the infeasibility of conventional decoding order strategy for secure communication among untrusted users. Then, a novel decoding order strategy is proposed that is efficient in securing all users' data from each other. 

\subsection{Conventional Decoding Order}
\begin{figure}[!t]
\centering
\includegraphics[scale=.4]{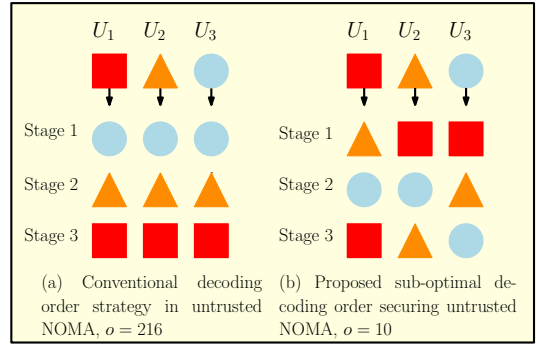}
\caption{Decoding order representation for 3-user untrusted NOMA system.}
\label{decoding_order_strategy}
\end{figure}

In the Conventional decoding order strategy of NOMA \cite{nomasurvey}, near users are considered to be trusted and far users never attempt to decode near users' data. Extending the conventional decoding order for untrusted NOMA case, where all the three users perform SIC based decoding in the order of weakest user to strongest user as shown in Fig. \ref{decoding_order_strategy} (a), decoding order can be represented as $\mathcal{D}_{o}=D_{216}=\{[3, 2, 1]^T,  [3, 2, 1]^T, [3, 2, 1]^T\}$. Here, we obtain $\mathbb{S}_{om}^{b3}=\{\phi\}$, $\mathbb{S}_{om}^{a3}=\{2, 1\}$; $\mathbb{S}_{om}^{b2}=\{3\}$, $\mathbb{S}_{om}^{a2}=\{1\}$; and $\mathbb{S}_{om}^{b1}=\{3, 2\}$, $\mathbb{S}_{om}^{a1}=\{\phi\}$ by $m$. As a result, $\gamma^{o}_{nm}$ defined in \eqref{SINR} can be expressed as 
\begin{equation}
\gamma^{o}_{nm}= \frac{\alpha_{n} |h_{m}|^2}{(\zeta\sum\limits_{i=n+1}^{3}\alpha_{i} + \sum\limits_{j=1}^{n-1}\alpha_{j})|h_{m}|^2+\frac{1}{\rho_{t}}}.
\end{equation}

Next, we present a result on inefficiency of conventional decoding order strategy in securing untrusted NOMA.
\begin{Lemma}
\textit{
Using conventional decoding order strategy in an untrusted NOMA scenario, the information of any weaker user cannot be secured from its respective stronger user.}
\end{Lemma}
\begin{IEEEproof}
In order to analyze the secrecy performance at weaker users, we investigate secrecy rate $R^{o}_{sn}$ at $U_{n}$ against $U_{m}$, where $m<n$. Using \eqref{secrecy_rate}, $R^{o}_{sn}$ can be written as
\begin{equation}\label{Rs2}
R^{o}_{sn} = \{\log_{2}(1+\gamma^{o}_{nn})- \underset{m\in\{1,2,...(n-1)\}}{\max} \log_{2}(1+\gamma^{o}_{nm})\}^{+}.
\end{equation}

As stated in Section \ref{section2}(C), the required condition for positive secrecy rate at $U_{n}$ is $\gamma^{o}_{nn}>\gamma^{o}_{nm}$, and this gives $|h_{n}|^{2}>| h_{m}|^{2}$ which is an infeasible condition as we have considered $|h_{m}|^{2}>| h_{n}|^{2}$. Thus, positive secrecy rate cannot be obtained at weaker user against stronger user.
\end{IEEEproof}
\begin{Remark}
 \textit{It can be easily inferred from the above study that only strongest user's information is safe from all others by using conventional NOMA strategy with untrusted users.} 
\end{Remark}

\subsection{Decoding Order Strategy from Secrecy Perspective}\label{Strategy}
Now, we present a decoding order strategy through which all feasible secure decoding orders can be obtained that are efficient in providing positive secrecy rate at all users.
\begin{Lemma}\label{lemma2}
\textit{In order to provide positive secrecy rate at any $U_{n}$ against any $U_{m}$ when $m<n$ in decoding order $\mathcal{D}_{o}$, the set of users decoded before decoding of $U_{n}$ by $U_{n}$, must not be equal to the set of users decoded before decoding of $U_{n}$ by $U_{m}$, i.e., $\mathbb{S}_{on}^{bn} \neq \mathbb{S}_{om}^{bn}$. }
\end{Lemma}
\begin{IEEEproof}
We prove this Lemma by contradiction. Let us suppose $\mathbb{S}_{on}^{bn} = \mathbb{S}_{om}^{bn}$. As a result, $\mathbb{S}_{on}^{an} = \mathbb{S}_{om}^{an}$. For analyzing the secrecy performance of $U_{n}$ against $U_{m}$, we observe that $\gamma^{o}_{nm}$ defined in \eqref{SINR}, when $U_{n}$ is decoded by $U_{m}$, can be given as
\begin{equation}
\gamma^{o}_{nm}= \frac{\alpha_{n} |h_{m}|^2}{(\zeta\sum\limits_{ i\in \mathbb{S}_{om}^{bn} }\alpha_{i}+\sum\limits_{ j\in \mathbb{S}_{om}^{an}}\alpha_{j})|h_{m}|^2+\frac{1}{\rho_{t}}}. 
\end{equation}

Similarly, in the case of decoding of $U_{n}$ by $U_{n}$, $\gamma^{o}_{nn}$ can be written as
\begin{equation}
\gamma^{o}_{nn}= \frac{\alpha_{n} |h_{n}|^2}{(\zeta\sum\limits_{ i\in \mathbb{S}_{on}^{bn} }\alpha_{i}+\sum\limits_{ j\in \mathbb{S}_{on}^{an}}\alpha_{j})|h_{n}|^2+\frac{1}{\rho_{t}}}. 
\end{equation}

Note that, due to the assumption of $\mathbb{S}_{on}^{bn} = \mathbb{S}_{om}^{bn}$, the required condition $\gamma^{o}_{nn}>\gamma^{o}_{nm}$ for positive secrecy at $U_{n}$ gives $|h_{n}|^{2}>| h_{m}|^{2}$ which is infeasible as $|h_{m}|^{2}>| h_{n}|^{2}$. Hence, it can be concluded that the decoding orders for which $\mathbb{S}_{on}^{bn} = \mathbb{S}_{om}^{bn}$, are not feasible to ensure secrecy to all users. 
\end{IEEEproof}

Based on the decoding order strategy proposed in Lemma $2$, now we present an algorithm that achieves the complete set $\mathbb{S}$ of \textit{secure decoding orders} that provide secrecy to all users. The detailed steps are shown in Algorithm \ref{Algo:AL1} given below. 
\begin{algorithm}[!htb]
{\small
\caption{\small Finding feasible secure decoding orders for untrusted NOMA through proposed decoding order strategy in Lemma $2$.}\label{Algo:AL1}
\begin{algorithmic}[1]
\Require Set $\mathbb{D}$ \hspace{2.9cm} $\triangleright$ set of total decoding orders
\Ensure Set $\mathbb{S}$ \hspace{2.8cm} $\triangleright$ set of secure decoding orders
\State Initialize $row=3$, $column=3$
\State Set counter to $0$
\For {each $\mathcal{D}_{o}$ in Set $\mathbb{D}$}
\State Initialize $flag1=1$
\For{$c1=2$ to $column$}   
\State Set $q=c1$
\For{$c2=1$ to $(c1-1)$} 
\For{$k=1$ to $row$}  
\If{($k=1$ $||$ $k=3$)}
\If{($\mathcal{D}_{o}(k,c1)=q$) $\&\&$ ($\mathcal{D}_{o}(k,c2)=q$)} 
\State Set $flag1=0$
\State break
\EndIf
\EndIf
\If{($k=2$)}
\If{($\mathcal{D}_{o}(k,c1)=q$) $\&\&$ ($\mathcal{D}_{o}(k,c2)=q$) $\&\&$ ($\mathcal{D}_{o}(k-1,c1)=\mathcal{D}_{o}(k-1,c2)$)}
 \State Set $flag1 = 0$
 \State break
\EndIf
\EndIf
\If{($flag1=0$)}
 \State break
 \EndIf
\EndFor
\If{($flag1=0$)}
\State break
\EndIf
\EndFor
\If{($flag1=0$)}
\State break
\EndIf
\EndFor
\If{($flag1=1$)}
\State  Increment counter
\State return $\mathbb{S}=\{\mathcal{D}_{o}\}$ \hspace{1cm} $\triangleright$ $\mathcal{D}_{o}$ as secure decoding order
\EndIf 
\EndFor
\end{algorithmic}
}
\end{algorithm}

\section{Sum Secrecy Rate Maximization}
In this section, we focus on maximizing the sum secrecy rate of the system by finding an appropriate secure decoding order and BS transmission PA to users.
\subsection{Problem Formulation}
Using our proposed decoding order strategy described in Lemma $2$, we obtain a set $\mathbb{S}$ of $76$ secure decoding orders out of $216$ total possible decoding orders, which
can ensure positive secrecy rate to all users. For a secure decoding order $\mathcal{D}_{o}$, where $\mathcal{D}_{o} \in \mathbb{S}$, the sum secrecy rate of the system, using \eqref{secrecy_rate}, can be defined as 
\begin{equation}\label{sumsecrecy}
\mathcal{R}^{o}_{s} = \sum_{n\in \mathcal{N}}R^{o}_{sn}.
\end{equation}

Main objective is to find an appropriate secure decoding order and PAs to users that can maximize sum secrecy rate of the system. Regarding this, the joint optimization problem over set $\mathbb{S}$ of secure decoding orders and BS transmission PA to users, using $\mathcal{R}^{o}_{s}$ \eqref{sumsecrecy}, can be formulated as
\begin{align}
&(P1): \underset{\mathcal{D}_{o}, \alpha_{n}, \mathcal{D}_{o} \in \mathbb{S}}{\text{maximize}} \quad \mathcal{R}^{o}_{s}, \nonumber 
\\ \text{subject to } \quad &(C1): 0 < \alpha_{n} < 1,  \quad
(C2): \sum_{n\in \mathcal{N}} \alpha_{n} = 1. \nonumber
\end{align}

The problem maximizing sum secrecy rate can be reformulated via considering the PA constraints $(C1)$ and $(C2)$, which can be simplified in terms of $\alpha_{1}$ and $\alpha_{2}$, and $\alpha_{3}=1-\alpha_{1}-\alpha_{2}$. Thus, the reformulated optimization problem can be stated as
\begin{align}
(P2): \underset{\mathcal{D}_{o}, \alpha_{1}, \alpha_{2}, \mathcal{D}_{o} \in \mathbb{S}}{\text{maximize}} \quad \mathcal{R}^{o}_{s}, \quad \text{subject to} \quad &(C3): \alpha_{1}+\alpha_{2} < 1, \nonumber
\\(C4): 0 < \alpha_{1}< 1, \quad &(C5): 0 < \alpha_{2}< 1. \nonumber
\end{align}

To solve the aforementioned joint optimization problem, we need to find feasible conditions on PAs for each secure decoding order in set $\mathbb{S}$. To reduce this computational complexity in finding an optimal solution, next, we propose an efficient sub-optimal decoding order solution. 

\subsection{Proposed Sub-optimal Decoding Order Solution}
In order to increase the secrecy rate at $U_{n}$ which further increases the sum secrecy rate of the system, $R_{nn}$ and $R_{nm}$ should be increased and decreased, respectively. This is because the required condition for positive secrecy rate at $U_{n}$, when $U_{n}$ is decoded by $U_{m}$, is $R_{nn}>R_{nm}$.  Based on these observations, we present two key insights in designing a sub-optimal decoding order as described in the following:
\\\textit{(i) $U_{n}$ decodes data of itself at the last stage:} To increase secrecy rate $\mathcal{R}^{o}_{sn}$ at $U_{n}$, $R_{nn}$ should be high. Note that the maximum $R_{nn}$ is obtained when the interference from all other users is maximum cancelled, which is attained when $U_{n}$ decodes its own data at the last stage. Hence, each user must decode its information after decoding all others' data. 
\\\textit{(ii) For the first $2$ stages, the data of other users are decoded by $U_{n}$, in accordance to one by one from the user with the strongest channel condition to the user with the weakest channel condition:} To maximize secrecy rate at $U_{n}$ against $U_{m}$, $R_{nm}$ should be decreased, which is obtained when interference from other users at $U_{m}$, while decoding data of $U_{n}$, is more. Thus, with a focus on improving the secrecy
rate for stronger users, the order of decoding of data from strongest to weakest channel helps in improving the overall secrecy rate of the system. Taking these points into account, we obtain a sub-optimal decoding order described as $\mathcal{D}_{10}$ in Fig. \ref{decoding_order_strategy}(b).

Now, we aim to find a joint-optimal solution of proposed optimization problem $(P2)$ for which we firstly solve the sum secrecy rate maximization problem for each decoding order.
\subsection{PA for each Secure Decoding Order}\label{PA}
The sum secrecy rate maximization problem for each secure decoding order $\mathcal{D}_{o}$ subject to PAs can be formulated as
\begin{equation}\label{problem_formulation1}
(P3): \underset{\alpha_{1}, \alpha_{2}}{\text{maximize}} \quad \mathcal{R}^{o}_{s}, \quad \text{subject to} \quad (C3), (C4), (C5). \nonumber
\end{equation}

Observing \eqref{SINR}-\eqref{secrecy_rate}, we note that the secrecy rate of each user is a non-convex function of PA parameters. In this case, we solve the sum secrecy rate maximization problem numerically and find BS transmission PA solutions $\alpha_{1}^{o*}$ and $\alpha_{2}^{o*}$ for $\mathcal{D}_{o}$. 

\subsection{Joint Solution of Decoding Order and PA}
As mentioned in Section \ref{PA}, first we numerically obtain optimized PAs and respective sum secrecy rate solution for all secure decoding orders of set $\mathbb{S}$. Let us define a set $\mathbb{R}=\{\mathcal{R}^{o*}_{s}\}$, where $\mathcal{R}^{o*}_{s}=\mathcal{R}^{o}_{s}(\alpha_{1}^{o*}, \alpha_{2}^{o*})$ denotes optimized sum secrecy rate of decoding order $\mathcal{D}_{o}$. Now, to complete joint optimization, we select the decoding order providing the maximum sum secrecy rate which can be expressed as
\begin{align}\label{joint1}
\hat{ \mathcal{D}_{o}} = \underset{\mathcal{D}_{o}\in \mathbb{S}}{\text{argmax}}(\mathcal{R}^{o*}_{s}). 
\end{align} 
The optimized sum secrecy rate of system can be given as  
\begin{align}\label{joint}
\hat{\mathcal{R}_{s}}(\hat\alpha_{1},\hat\alpha_{2})= \underset{\mathcal{D}_{o}\in \mathbb{S}} \max(\mathbb{R}), 
\end{align}
where $(\hat\alpha_{1}$, $\hat\alpha_{2})$ denotes numerically optimized system PAs. 
Note that, the computational complexity to obtain optimized result is $76$ times more as compared to sub-optimal approach.
\section{Numerical Results and Discussion}
Now, numerical results are presented to assess the performance of proposed design under various system settings. For simulation, $L_{p}=1$, $e = 3$ and $\zeta= 0.1$ are considered. Noise signal at all users follow Gaussian distribution with a noise power of $-90$ dBm. Symbolic PA $\alpha_{n}=\frac{1}{{(|h_{n}|^2)^\beta} (\sum_{ n\in \mathcal{N}} \frac{1}{(|h_{n}|^2)^{\beta}})}$ is considered, indicating PA to users is in accordance with users' channel conditions. Note that $\beta$ is a coefficient ranging from $-1$ to $1$. Simulation results are averaged over $10^4$ randomly generated channel realizations.

\subsection{Insights on Decoding Order Selection}
\begin{figure}[!t]
\centering
\includegraphics[scale=.37]{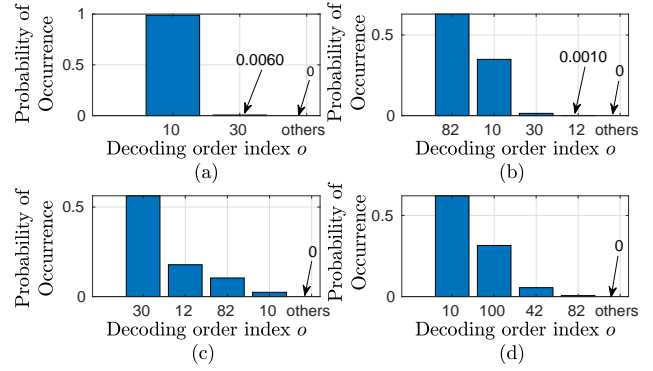}
\caption{Probability of occurrence of different secure decoding orders with maximum sum secrecy rate at various PAs, (a) $\alpha_{1}=0.3333$, $\alpha_{2}=0.3333$, (b) $\alpha_{1}=0.1667$, $\alpha_{2}=0.3333$, (c) $\beta=1$, and (d) $\beta=-1$.}
\label{plot1}
\end{figure}

Fig. \ref{plot1} presents the probability of occurrence of secure decoding orders with a maximum sum secrecy rate for various channel realizations. Fig. \ref{plot1}(a) and Fig. \ref{plot1}(b) are plotted for fixed PAs to all users for all  channel realizations and Fig. \ref{plot1}(c) and Fig. \ref{plot1}(d) are plotted, respectively, for $\beta=1$ and $\beta=-1$. We observe that different decoding orders win for various PA schemes through which we can conclude that an appropriate decoding order plays a vital role in designing an optimal secure communication system.

\subsection{Insights on Optimality}
\begin{figure}[!t]
\centering
\includegraphics[scale=.37]{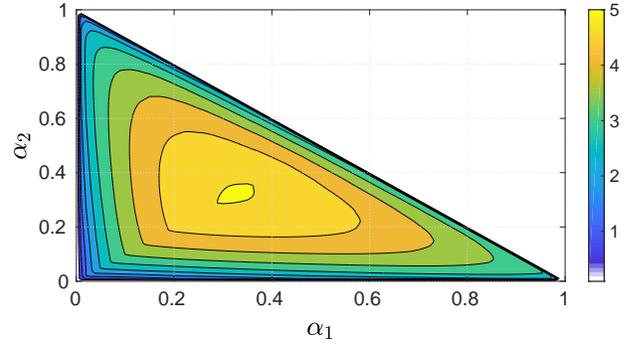}
\caption{Insight on optimality in sub-optimal decoding order $o= 10$.}
\label{plot2}
\end{figure}

Now we present the numerical proof that the unique solution of PA exists for each secure decoding order $\mathcal{D}_{o}$ of set $\mathbb{S}$. In this context, Fig. \ref{plot2} plots sum secrecy rate performance of the system for various PAs in a sub-optimal decoding order $\mathcal{D}_{10}$. One can easily observe that the contour plot in Fig. \ref{plot2} confirms existence of optimal PAs to users achieving maximum sum secrecy rate. Thus, it can be concluded that the optimal  secrecy rate performance of a secure NOMA system is highly dependent on BS transmit PA to users.

\subsection{Insights on Sub-optimal Decoding Order Performance}
\begin{figure}[!t]
\centering
\includegraphics[scale=.37]{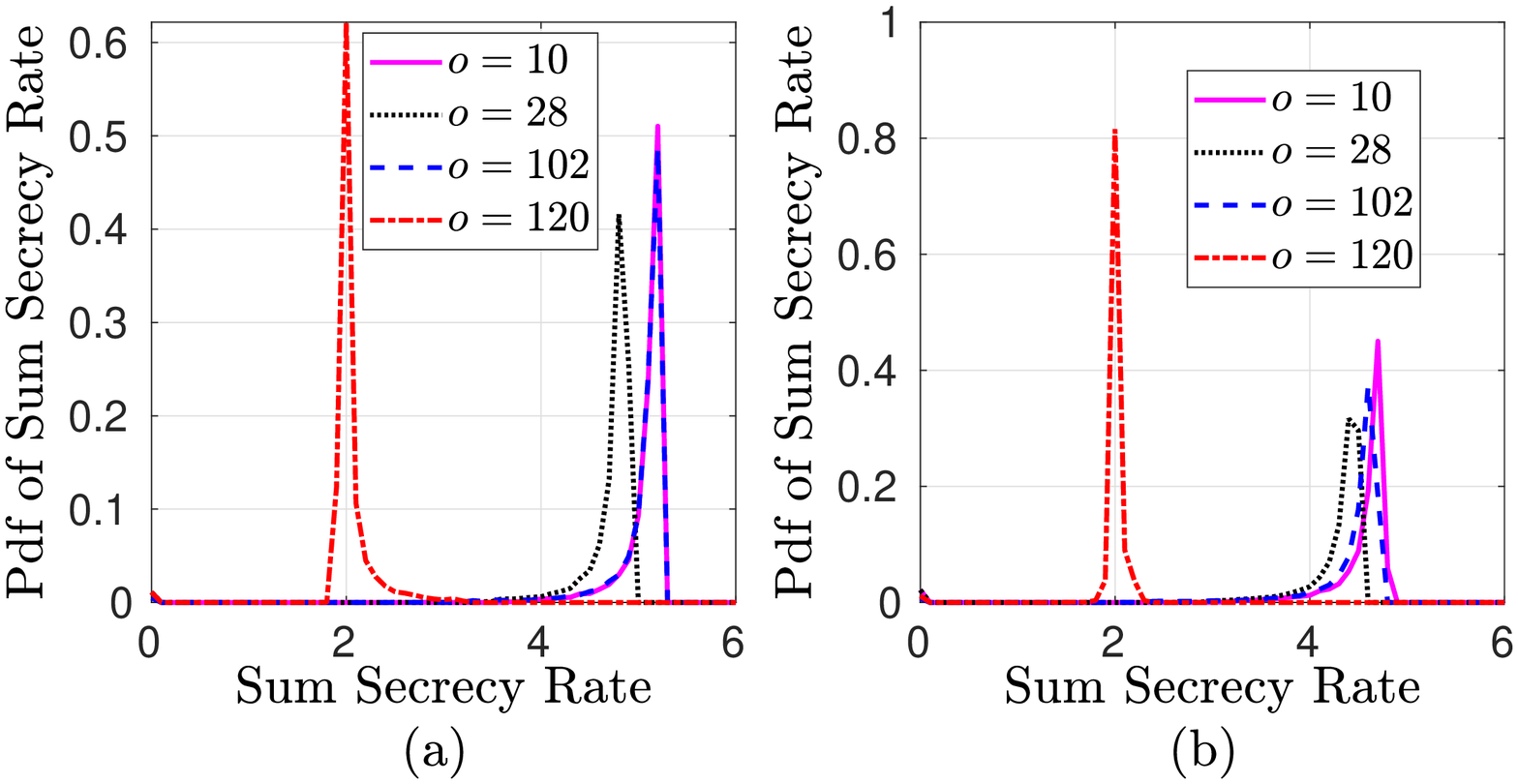}
\caption{Impact of different secure decoding orders on the average sum secrecy rate performance of the system at fixed PAs, (a) $\alpha_{1}=0.3333$, $\alpha_{2}=0.3333$, (b) $\alpha_{1}=0.1667$, $\alpha_{2}=0.3333$.}
\label{plot3}
\end{figure}

Fig. \ref{plot3} is plotted to present the impact of the proposed sub-optimal decoding order over other decoding orders.
We plot the probability density function (Pdf) of the system's sum secrecy rate for $4$ different secure decoding orders with two different PAs to users. Observing the variance of the Pdf plots, it can be concluded that the proposed sub-optimal decoding order outperforms the remaining secure decoding orders. 

\subsection{Impact of Relative Distance between Users}
\begin{figure}[!t]
\centering
\includegraphics[scale=.37]{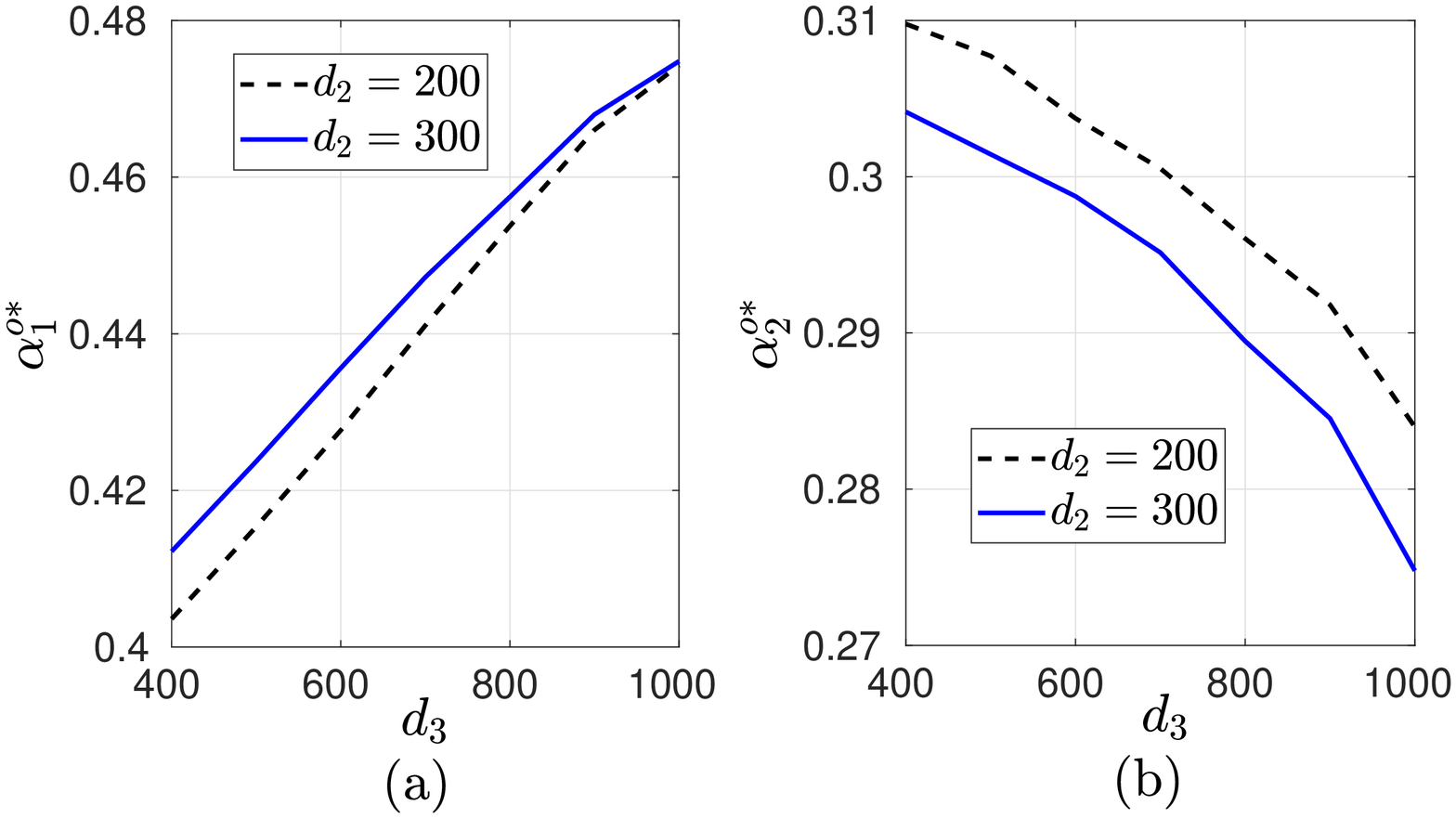}
\caption{Variation of optimized PAs with users' distance for a decoding order index $o= 100$,  $d_{1}= 100$ meter.}
\label{plot4}
\end{figure}

Here Fig. \ref{plot4} present the impact of users' distance on optimized PAs performance that maximize the sum secrecy rate of the system. Fixing distance $d_{1}=100$ meter, the effect of variation of $d_{2}$ and $d_{3}$ from BS is observed on optimized PAs for a decoding order index $o=100$. The results presented in Fig. \ref{plot4}(a) show that $\alpha_{1}^{o*}$ monotonically increases with the increase in  $d_{2}$ and $d_{3}$. On the other hand, increasing $d_{2}$ and $d_{3}$ decrease achievable $\alpha_{2}^{o*}$ as shown in Fig. \ref{plot4}(b). The reason is, with increasing distance of weaker users, a decrease in achievable information rate at weaker users improves secrecy rate for stronger users. In this case, decreasing PAs to weaker users helps in further increasing the secrecy rate at respective stronger users, and thus, the optimal sum secrecy rate obtains.

\subsection{Performance Comparison}
\begin{figure}[!t]
\centering
\includegraphics[scale=.37]{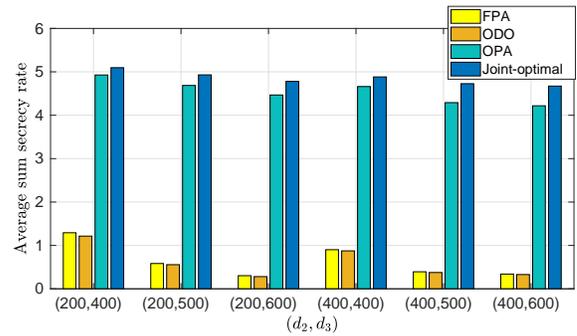}
\caption{Performance comparison of jointly optimized scheme with FPA, ODO, and OPA schemes, $d_{1}=100$ meter.}
\label{plot5}
\end{figure}

To demonstrate the performance gain achieved by the joint-optimal scheme (joint-optimal decoding order and PAs to users) for sum secrecy rate maximization, Fig. \ref{plot5} presents its performance comparison with
three different schemes (FPA: fixed PA for sub-optimal decoding order; ODO: fixed PA for optimal decoding order; OPA: optimal PA for sub-optimal decoding order). Note that, fixed PAs are considered as $\alpha_{1}=0.1667$ and $\alpha_{2}=0.3333$. The results depict that the jointly optimized solution achieves best sum secrecy rate performance and obtains average performance improvement of about $27$ dB. 

\section{Concluding Remarks}
By employing PLS to NOMA, we have focused on resolving the inherent security issue among untrusted NOMA users. A novel decoding order strategy has been proposed that is efficient in providing positive secrecy rate at all users. Based on the proposed decoding order strategy, numerically, we have obtained all feasible secure decoding orders. To maximize the sum secrecy rate of the system, we have investigated joint optimization over secure decoding orders and PAs. Also, a sub-optimal decoding order solution is proposed. Numerical results show that the optimal solution achieves average gain of about $27$ dB in the sum secrecy rate over relevant schemes.

\section*{Acknowledgement}
This work has been supported by the Tata Consultancy Services (TCS) Research Scholar Program Fellowship.

\bibliographystyle{IEEEtran}
\bibliography{ref}
\end{document}